\documentstyle[12pt]{article}

\begin{document}
\vskip 0.2cm
\hfill{YITP-SB-03-13}
\vskip 0.2cm
\hfill{INLO-PUB-04/03}\\[0.5cm]
\vskip 0.2cm
\centerline{\large\bf
{Higgs Production at NNLO} }
\vskip 0.4cm
\centerline {\sc V. Ravindran
\footnote {Talk given at PASCOS'03 at TIFR, Mumbai}}
\centerline{\it Harish-Chandra Research Institute,}
\centerline{\it Chhatnag Road, Jhunsi,}
\centerline{\it Allahabad, 211019, India.}
\vskip 0.2cm
\centerline {\sc J. Smith
}
\centerline{\it C.N. Yang Institute for Theoretical Physics,}
\centerline{\it State University of New York at Stony Brook,
New York 11794-3840, USA.}
\vskip 0.2cm
\centerline {\sc W.L. van Neerven}
\centerline{\it Instituut-Lorentz}
\centerline{\it University of Leiden,}
\centerline{\it PO Box 9506, 2300 RA Leiden,}
\centerline{\it The Netherlands.}
\vskip 0.2cm
\vskip 0.2cm
\centerline{\bf Abstract}
\vskip 0.3cm
We describe the calculation of inclusive Higgs boson production at hadronic
colliders at next-to-next-to-leading order (NNLO) in perturbative quantum
chromodynamics.  We have used the technique developed in reference [4].
Our results agree with those published earlier in the literature.

\section{Introduction}
The discovery of the Higgs boson will shed light on the
symmetry breaking machanism of the Standard Model (SM). The experimental
bound from the LEP experiments and precision studies within and beyond
the standard model strongly suggest that hadron colliders such as the Tevatron
and the LHC will see the Higgs boson if it exists.
At these machines the dominant contribution to single Higgs boson production
is the gluon-gluon fusion process through heavy quark loops.
The reason for this is that the Higgs boson couples strongly to
heavy quarks.  In addition the large gluon flux at the LHC
enhances the total inclusive cross section substantially.
The NLO corrections along with NLO parton distributions yield a large $K$
factor and also show a strong scale dependence. Hence there is need for
improved parton densities using NNLO splitting functions as well as
inclusion of NNLO partonic cross sections.  The NNLO correction to Higgs boson
production was first computed by an expansion technique in \cite{kilgore}.
Exact results were obtained in \cite{anastasiou} using Cutkosky rules.
In our work \cite{rsn}, we have used a straightforward technique which
was adopted in \cite{vanNeerven1} to compute the NNLO corrections to
Drell-Yan process.  Clever choice of the integration variables
in specific frames makes the computation manageable.

\section{Method of Computation}
We use the effective Lagrangian approach which emerges
from the SM in the heavy top quark limit ($m_t \rightarrow \infty$) and
is found to be a good approximation at hadron colliders.
For Higgs boson production at hadron colliders at the NNLO level
one has to compute
1) tree level $a+b \rightarrow c+d+ H$,
2) one-loop corrected $a+b \rightarrow c+H$ and 3) two-loop corrected
$g+g \rightarrow H$,
where $a,b,c,d$ are light partons such as quarks, antiquarks and gluons
whose interactions are governed by QCD.  These corrections involve the
computation of $2 \rightarrow 3$ body phase-space integrals and two and
one-loop momentum integrals followed by
$2 \rightarrow 1$ and $2 \rightarrow 2$ phase-space
integrations respectively.  We use dimensional regularization
(space time dimension is taken to be $n=4+\varepsilon$) to regulate
both ultraviolet and infrared (soft and collinear) divergences.
We first describe here how we have performed three-body phase-space integrals
for $2 \rightarrow 3$ tree level matrix elements and
the two-body phase-space integrals for one-loop corrected matrix elements.
The $2 \rightarrow 3$ body processes involve two angular integrations
(say $\theta,\phi$) and two parametric integrations
(say $z,y$).  Before we perform these integrations, it is important to
classify the matrix elements in such a way that the phase-space integrations
over them can be done in suitable frames.  For example, when the Higgs
boson is produced from the incoming partons, the center of mass (CM)
frame of outgoing partons is the most suitable frame, because
the massive propagators $1/(P_{15}^\alpha P_{25}^\beta)$ (
$\alpha,\beta \geq 1$) will not involve angular dependence in this
frame. Here $P_{i5}=(p_i+p_H)^2$ where $p_H$ is the momentum of the Higgs
boson, and $p_i$ is the momentum of the massless parton.
Similarly when the Higgs boson is produced from an outgoing parton, we choose
the CM frame of incoming partons where $1/(P_{35}^\alpha P_{45}^\beta)$
($\alpha,\beta \geq 1$) do not depend on the angles.  Complications arise
when we encounter processes where the Higgs boson is produced by both
initial and final state partons, i.e., where interference terms of the form
$1/(P_{15}^\alpha P_{45}^\beta)$ ($\alpha,\beta \geq 0$) appear. In this case
we have chosen the CM frame of the 4th (or 3rd) parton and the Higgs boson
where the angular integrals and other parametric integrals are less difficult.
In the CM frame of the incoming partons and the CM frame of the
outgoing partons we perform the angular
integrations exactly using the result given in \cite{vanNeerven2}.
\begin{eqnarray}
&&\int_0^\pi d \theta \int_0^\pi d\phi ~~
\sin^{n-3}\theta \sin^{n-4}\phi \, {\cal C}_{ij}(\theta,\phi,\chi)=
2^{1-i-j} \pi
\nonumber\\ &&
\times {\Gamma({1 \over 2}n-1-j)\Gamma({1 \over 2}n-1-i)
\Gamma(n-3) \over
\Gamma(n-2-i-j)
\Gamma^2({1 \over 2}n-1) }
F_{1,2}\Big(i,j,{1\over 2} n-1;\cos^2(\chi/2)\Big)\,,
\nonumber
\end{eqnarray}
where ${\cal C}_{ij}=(1-\cos\theta)^{-i} (1-\cos\chi\cos\theta
-\sin\chi \cos\phi \sin\theta)^{-j}$. Here $\cos\chi$ is
related to kinematical variables such as $x~(=m_H^2/s$, $s$-CM energy) and
the integration variables $z$ and $y$, and $F_{1,2}$ is the hypergeometric
function. In the CM frame of the 4th parton and Higgs boson, due
to the complexity, we performed one angular integration (say $\phi$)
exactly and performed the remaining $\theta$ integration after
expanding the integrands in powers of $\varepsilon=n-4$.
In all these frames, using various Kummer's relations,
the hypergeometric functions are simplified to the form
$F_{1,2}(\pm\varepsilon/2,\pm\varepsilon/2,1\pm\varepsilon/2;{\cal D}(x,y,z))$
which is the most suitable for integrations over $z$ and $y$.
The next hurdle in the computation is the appearance of terms with
large powers in $1/(1-z)$ or $1/z$.
We have reduced the higher powers of $1/(1-z)^{\alpha+\beta\varepsilon}$
or $1/z^{\alpha+\beta\varepsilon}$
where $\alpha > 1$ by successive integration by parts with exact
hypergeometric functions until we arrive at $1/(1-z)^{1+\beta \varepsilon}$ or
$1/z^{1+\beta\varepsilon}$ multiplied by regular functions.
We have used the following identity to accomplish this:
\begin{eqnarray}
{d \over dz}F_{1,2}({\varepsilon \over 2},{\varepsilon \over 2},1+{\varepsilon \over 2};{\cal D}(z))&=&
{\varepsilon \over 2 {\cal D}(z)} {d {\cal D}(z)\over dz}\Big(\big(1
-{\cal D}(z)\big)^{-{\varepsilon \over 2}}
\nonumber\\
&&
-F_{1,2}\big({\varepsilon \over 2},{\varepsilon \over 2},1+{\varepsilon \over 2};{\cal D}(z)\big)\Big)
\,.
\end{eqnarray}
In the end we are left with integrations of
the form $\int_0^1 dz z^{-1-\beta \varepsilon}
f(z)$ and/or $\int_0^1 dz (1-z)^{-1-\beta \varepsilon}
f(z)$.  Such integrals are simplified as follows:
\begin{eqnarray}
\int_0^1 dz  z^{-1-\beta \varepsilon}
f(z) = \int_0^\delta dz ~z^{-1-\beta \varepsilon} f(z)+
  \int_\delta^1 dz~ z^{-1-\beta \varepsilon} f(z) \quad \quad \delta << 1\,.
\end{eqnarray}
The first term can be evaluated to be
$f(0) ( [\beta \varepsilon]^{-1}+\log\delta+
[\beta \varepsilon / 2] \log^2\delta+\cdot \cdot \cdot)$.
After expanding $z^{-\beta \varepsilon} $ in powers
of $\varepsilon$ in the second term the $z$ integration can be
performed exactly order-by-order in $\varepsilon$ with non-zero $\delta$.
At the end the $\delta$ dependence cancels in each order in $\varepsilon$.
Since the $z$ integration over the hypergeometric functions
is nontrivial due to their complicated arguments, we have expanded them in
powers of $\varepsilon$ prior to the $z$ integration:
\begin{eqnarray}
F_{1,2}( {\varepsilon \over 2}, {\varepsilon \over 2},1+ {\varepsilon \over 2};{\cal D}(z))&=&
1 + {\varepsilon^2 \over 4} {\rm Li}_2({\cal D}(z))
+ {\varepsilon^3 \over 8} \Big({\rm S}_{1,2}({\cal D}(z))
- {\rm Li}_{3}({\cal D}(z))\Big)
\nonumber \\
&&
+{\varepsilon^4 \over 16} \Big({\rm S}_{1,3}({\cal D}(z))-{\rm S}_{2,2}
({\cal D}(z)) +{\rm Li}_4({\cal D}(z))\Big) \,,
\end{eqnarray}
where
${\rm S}_{n,p}(z)=(-1)^{n+p-1} [(n-1)! p!]^{-1}
\int_0^1 dt [t]^{-1} \log^{n-1}(t) \log^p(1-z t) $ with
$ n,p \geq 1$ and
${\rm Li}_n(z)={\rm S}_{n-1,1}(z)$(see \cite{lewin}).
We have repeated the same procedure to perform the remaining $y$ integration.
In addition to $2 \rightarrow 3$ contributions, we encounter
one-loop corrected $2 \rightarrow 2$ processes at NNLO level.
Here the one-loop tensorial integrals are reduced to scalar integrals
using the Passarino-Veltman reduction procedure implemented in $n$ dimensions.
The resulting one-loop two and three-point scalar integrals can be
expressed in terms of kinematic invariants.
In the case of the four-point function, the scalar integrals can be expressed
only in terms of hypergeometric functions which increases the complexity of
the two-body phase-space integrations.  We follow the procedure adopted
for the $2\rightarrow 3$ phase-space integrations to perform
the two-body phase-space integrations.  After performing all these integrals,
we have removed all the ultraviolet divergences by strong coupling
and operator renormalization constants.  The remaining collinear divergences
are removed by mass factorization. Then we are left with finite
partonic cross sections which are folded with parton distribution functions
to compute hadronic cross section for the inclusive Higgs boson production.

Using the method described above we have successfully computed the NNLO
corrections to Higgs boson production at hadron colliders and found complete
agreement with the results of \cite{kilgore,anastasiou}.
We find that NNLO corrections improve the convergence of the perturbative
result and decrease the scale ambiguities inherent in it.

\end{document}